# A Framework for Evaluating the Retrieval Effectiveness of Search Engines

**Dirk Lewandowski**
*Hamburg University of Applied Sciences, Germany*

## ABSTRACT

This chapter presents a theoretical framework for evaluating next generation search engines. We focus on search engines whose results presentation is enriched with additional information and does not merely present the usual list of "10 blue links", that is, of ten links to results, accompanied by a short description. While Web search is used as an example here, the framework can easily be applied to search engines in any other area.
The framework not only addresses the results presentation, but also takes into account an extension of the general design of retrieval effectiveness tests. The chapter examines the ways in which this design might influence the results of such studies and how a reliable test is best designed.

## INTRODUCTION

Information retrieval systems in general and specific search engines need to be evaluated during the development process, as well as when the system is running. A main objective of the evaluations is to improve the quality of the search results, although other reasons for evaluating search engines do exist (see Lewandowski & Höchstötter, 2008). A variety of quality factors can be applied to search engines. These can be grouped into four major areas (Lewandowski & Höchstötter, 2008):

- Index Quality: This area of quality measurement indicates the important role that search engines' databases play in retrieving relevant and comprehensive results. Areas of interest include Web coverage (e.g., Gulli & Signorini, 2005), country bias (e.g., Liwen Vaughan & Thelwall, 2004; Liwen Vaughan & Zhang, 2007), and freshness (e.g., Lewandowski, 2008a; Lewandowski, Wahlig, & Meyer-Bautor, 2006).
- Quality of the results: Derivates of classic retrieval tests are applied here. However, which measures should be applied and whether or not new measures are needed to satisfy the unique character of the search engines and their users should be considered (Lewandowski, 2008d).
- Quality of search features: A sufficient set of search features and a sophisticated query language should be offered and should function reliably (e.g., Lewandowski, 2004, 2008b).
- Search engine usability: The question is whether it is possible for users to interact with search engines in an efficient and effective way.

While all the areas mentioned are of great importance, this chapter will focus on ways in which to measure the quality of the search results, a central aspect of search engine evaluation. Nonetheless, it is imperative to realize that a search engine that offers perfect results may still not be accepted by its users, due, for example, to usability failures.

This chapter will describe a framework for evaluating next generation search engines, whether they are Web search engines or more specific applications. A search engine in the context of this chapter refers to an information retrieval system that searches a considerably large database of unstructured or semi-structured data (as opposed to a general information retrieval system that searches a structured database). A *next-generation search engine* is a search engine that does not present its results as a simple list, but makes use of advanced forms of results presentation; that is to say, they enrich the list-based results presentation with additional information or the results are presented in a different style. Thus, the results are unequally presented in terms of *screen real estate*, that is, in terms of the area on the results screen that each results description is granted.

While the term "search engine" is often equated with "Web search engine," in this chapter, all types of search engines are considered, although points regarding Web search engines are particularly emphasized, as they constitute the major area of our research.

The remainder of this chapter is organized as follows: First, a certain amount of background information on information retrieval evaluation is presented; next, the relevant literature related to such areas of interest as search engine retrieval effectiveness tests, click-through analysis, search engine user behavior, and results presentation. Then, a framework for search engine retrieval effectiveness evaluation, which will be described in detail, is presented. The chapter concludes by alluding to future research directions and with a set of concluding remarks.

## BACKGROUND

Evaluation has always been an important aspect (and an important research area) of information retrieval. Most studies follow (at least to a certain degree) the Cranfield paradigm, using a set of ad-hoc queries for evaluation and calculating effectiveness measures such as precision and recall. While the Cranfield paradigm has often been criticized, it is not without merit and is used in large evaluation initiatives such as TREC and CLEF, which were designed to evaluate search results. Most search engine evaluations today are "TREC-style,"[1] as they follow the approach used in these tests.

They use, however, a somewhat limited understanding of a user's behavior. As will become apparent in the literature review section, user results are determinant upon selection behavior, which is influenced by many factors. However, TREC-style evaluations focus on a "dedicated searcher," i.e., someone who is willing to examine every result given by the search engine and follow the exact order in which the results are presented (Harman & Voorhees, 2006, p. 117). Additionally, these evaluations assume that a user is interested in a high recall, as well as in a high level of precision of the results. Finally, the user is willing to consider a large number of documents for his query (Harman & Voorhees, 2006, p. 117). It can be easily seen that these assumptions may not hold true for each use scenario; for example, in Web searches, users usually only examine a few results presented at the top of the results screen, and not necessarily in the order in which they are presented. Web search engine users are usually interested in a few good-quality documents (a high precision, at least on the first few results positions) and not interested in a complete set of all relevant results (a high recall).

However, even if the kind of searching behavior assumed in TREC-style evaluations were once accurate in other contexts than Web search, now it is no longer applicable. As Web search engines largely influence user behavior in all searching contexts, there is a need for new models for evaluating all types of search engines.

## LITERATURE REVIEW

In this section, we review the literature on retrieval effectiveness tests with a focus on Web search engines. We will see that most of the problems that occur in the evaluation of Web search engines also occur when the subject of evaluation is an entirely different kind of search engine. In addition, we will briefly review the literature on search engine user behavior and on results presentation in search engines. From the literature review, we will derive the major points that will be implemented in our search engine evaluation framework, presented later.

### Retrieval effectiveness tests

Retrieval effectiveness tests should focus on only one query type (either informational, navigational, or transactional queries – for a detailed discussion on query types, see below). While for informational queries, a results *set* can be considered, navigational queries are usually satisfied with only one result. For the satisfaction of transactional queries, a certain degree of interaction is needed. To measure the performance of a search engine on such queries, these interactions must be modeled. Therefore, simple retrieval effectiveness tests are not sufficient; one should use a combined user- and effectiveness study.

In our review, only studies using informational queries (cf. Broder, 2002) will be considered. While there are some existing studies on the performance of Web search engines on navigational queries (see Hawking & Craswell, 2005; Lewandowski, 2011a), these should be considered separately because of the very different information needs being considered. However, there are some "mixed studies" (for example, Griesbaum, 2004) that use query sets consisting of informational as well as navigational queries. These studies are flawed because, in contrast to informational queries, the expected result for a navigational query is just one result. When a greater number of results are considered, even a search engine that found the desired page and placed it in the first position would receive bad precision values. Therefore, informational and navigational queries should be strictly separated.

In general, retrieval effectiveness tests rely on methods derived from the design of information retrieval tests (Tague-Sucliffe, 1992) and on advice that focuses more on Web search engines (Gordon & Pathak, 1999; Hawking, Craswell, Bailey, & Griffiths, 2001). However, there are some remarkable differences in study design, which are reviewed in the following paragraphs.

First, the number of queries used in the studies varies greatly. The oldest studies, especially (Chu & Rosenthal, 1996; Ding & Marchionini, 1996; Leighton & Srivastava, 1999), use only a few queries (5 to 15) and are, therefore, of limited use (for a discussion of the minimum number of queries that should be used in such tests, see Buckley & Voorhees, 2000). Newer studies use a minimum of 25 queries, some 50 or more.

In older studies, queries are usually taken from reference questions or commercial online systems, while newer studies focus more on the general users' interests or combine both types of questions. There are studies that deal with a special set of query topics (for example, business; see Gordon & Pathak, 1999), but a trend is observable in focusing on the general user in search engine testing (cf. Lewandowski, 2008d).

Regarding the number of results taken into account, most investigations consider only the first 10 or 20 results. This is due not only to the amount of work required for the evaluators but also to the general behavior of search engine users. These users only seldom view more than the first results page. While these results pages typically consist of 10 results (the infamous "ten blue links"), search engines now tend to add additional results (from news or video databases, for example), so that the total number of results, especially on the first results page, is often considerably higher (Höchstötter & Lewandowski, 2009).

Furthermore, researchers found that general Web search engine users heavily focus on the first few results (Cutrell & Guan, 2007; Granka, Hembrooke, & Gay, 2005; Hotchkiss, 2007; Joachims, Granka, Pan, Hembrooke, & Gay, 2005), which are shown in the so-called "visible area" (Höchstötter & Lewandowski, 2009)— that is, the section of the first search engine results page (SERP) that is visible without scrolling down. Keeping these findings in mind, a cut-off value of 10 in search engine retrieval effectiveness tests seems reasonable.

An important question is how the results should be judged. Most studies use relevance scales (with three to six points). Griesbaum's studies (Griesbaum, 2004; Griesbaum, Rittberger, & Bekavac, 2002) use binary relevance judgments with one exception: results can also be judged as "pointing to a relevant document" (for example, the page itself is not relevant but contains a hyperlink to a relevant page). This is so as to take into account the special nature of the Web. However, it seems problematic to judge these pages as somewhat relevant, as pages may contain dozens or even hundreds of links, and a user would then (in bad circumstances) have to follow a number of links to access the relevant document.

Two particularly important points in the evaluation of search engine results are whether the source of the results is made anonymous (i.e., the jurors do not know which search engine delivered a certain result) and whether the results lists are randomized. Randomization is important in order to avoid learning effects. While most of the newer studies anonymize the results (as far as which search engine produced the results), only three studies were found that randomized the results lists (Gordon & Pathak, 1999; Lewandowski, 2008d; Véronis, 2006).

Most studies make use of students as jurors. This comes as no surprise, as most researchers teach courses where they have access to students that can serve as jurors. In some cases (again, mainly older studies), the researchers themselves are the ones to judge the documents (Chu & Rosenthal, 1996; Ding & Marchionini, 1996; Dresel et al., 2001; Griesbaum, 2004; Griesbaum et al., 2002; Leighton & Srivastava, 1999).

To our knowledge, no study regarding the effectiveness of search engine retrieval utilizes multiple jurors, a fact that may result in flawed results, as different jurors may not agree on the relevance of individual results.

While retrieval effectiveness tests that make use of jurors are time-consuming and thus expensive, a cheaper approach involves using click-through data to determine the relevance of certain results. The quality of a result is determined by the number of times it was selected on the results page, the length of time users spent reading the result ("dwell time"), and the number of times a user directly returned to the results screen after screening the results clicked ("bounce rate"). While such studies have the advantage of considering a large number of queries and results (e.g., Joachims et al., 2005), one significant disadvantage is that only results that users actually clicked can be considered. As discussed in the following section, users only view a small portion of the results list, and click-through data is only available for this portion. Utilizing click-through data to improve the performance of a search engine may lead to a "rich get richer" effect, as results presented at the top of the results list are clicked more often than those positioned lower on the list.

Lewandowski (2008d) provides an overview of the major Web search engine retrieval effectiveness tests conducted between 1996 and 2007. For an overview of older tests, see Gordon & Pathak (1999, p. 148).

## Results presentation

Results presentation in Web search engines has dramatically changed from a list presentation of "10 blue links", that is, ten results, each equally presented with a title, description, and URL, to individual descriptions (containing additional information for certain results), different sizes, and graphically supported presentations of certain results. While findings regarding the composition of search engine results pages with a focus on Web search engines are presented here, similar results presentations can also be found in other contexts. It seems likely that even more results pages composed from different sources ("Universal Search") will make an appearance in the future, as users' expectations are largely guided by what they see in their daily use of Web search engines.

Table 1 provides an overview of the different types of results that can be presented on search engine results pages. An example of an extended presentation of search engine results is given in Figure 1.

| Name | Description | Position |
|------|-------------|----------|
| Organic | Results from Web crawl. "Objective hits" not influenced by direct payments. | Central on results page. |
| Sponsored | Paid results, separated from the organic results list. | Above or below organic results, on the right-hand side of the results list. |
| Shortcuts | Emphasized results, pointing to results from a special collection. | Above organic results, within organic results list. |
| Primary search result | Extended result that highlights different collections and is accompanied by an image as well as further information. | Above organic results, often within organic results. |
| Prefetch | Results from a preferred source, emphasized in the results set. | Above or within organic results. |
| Snippet | Regular organic result, for which the result description is extended by additional navigational links. | Within organic results list (usually only in first position). |
| Child | Second result from the same server, with a link to further results from the same server. | Within organic results list; indented. |

*Table 1. Typical elements presented on Web search engine results pages (Höchstötter & Lewandowski, 2009).*

*Figure 1. Results presentation (complete SERP) in the Google search engine (Höchstötter & Lewandowski, 2009, p. 1798).*

While search engines usually return thousands of results, users are not willing to view more than a few (Jansen & Spink, 2006; Keane, O'Brien, & Smyth, 2008; Lorigo et al., 2008), which is the reason why search engines, for all practical purposes, have only the first page to present relevant results. Even in cases in which users are willing to view more than the first few results, the initial ranking heavily influences their perception of the results set (that is, when the first few results are not considered relevant, the query is modified or the search is abandoned).

This first results page must also be used to present results from additional collections, as users usually do not follow links to these additional collections, the so-called "tabs". Search engine expert Danny Sullivan even coined the term "tab blindness" for this phenomenon (Sullivan, 2003).

Results beyond organic results also require space on the results page. With the addition of more and more results from special collections (and, in some cases, the addition of more and more ads above the organic results), a general change in results presentation can be observed (Lewandowski, 2008c). Organic results become less important, as additional results take their place. Nicholson et al. (2006) calculate the "editorial precision" (EPrec) of SERPs by dividing the total screen space by the space used for editorial (that is, organic) results. While this differentiation is mainly useful for Web search engine results pages, editorial precision can be calculated for any element of the results page and should be considered in retrieval tests.

## Search engine user behavior

Findings on search engine user behavior indicate that users are not willing to spend much time and cognitive resources on formulating search queries (Machill, Neuberger, Schweiger, & Wirth, 2004), a fact that results in short, unspecific queries (e.g., Höchstötter & Koch, 2008; Jansen & Spink, 2006) . While this surely applies to Web searching, similar behavior can also be found in other contexts, such as in scientific searching (Rowlands et al., 2008) or library searches (Hennies & Dressler, 2006).

Search queries tend to be very short and do not show variations over a longitudinal period. Nearly half of the search queries in Web searching still only contain one term. On average, a query contains between 1.6 and 3.3 terms, depending on the query language (Höchstötter & Koch, 2008; Jansen & Spink, 2006). Höchstötter and Koch (2008) give an overview of different studies measuring users' querying behavior (including query length and complexity).

Most searching persons evaluate the results listings very quickly before clicking on one or two recommended Web pages (Hotchkiss, Garrison, & Jensen, 2004; Spink & Jansen, 2004). Users consider only some of the results provided, mainly those results presented at the top of the ranked lists (Granka, Joachims, & Gay, 2004; Pan et al., 2007), and even more prefer the results presented in the "visible area" of the results screens, that is to say, the results visible without scrolling down the page (Höchstötter & Lewandowski, 2009). Lastly, results selection is determined by presenting some results in a different manner than the other results (see Höchstötter & Lewandowski, 2009), that is, emphasizing certain results by means of the use of color, frames, or size. Figure 2 gives an example of a results presentation that emphasizes a certain result with a picture and an enriched results description.

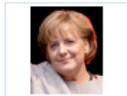

*Figure 2. Results presentation (clipping from a Yahoo results list), emphasizing a certain result by using a picture.*

Selection behavior from the search-engine results lists explicitly indicate just how much users rely on the engine-based ranking (Granka , Joachims, & Gay, 2004; Joachims, Granka, Pan, Hembrooke, & Gay, 2005; Loriga et al., 2008). Not only do a significant number of users limit themselves to screening the first page of results (Höchstötter & Koch, 2008), they also focus heavily on the top results.

Eye-tracking studies (e.g., Cutrell & Guan, 2007; Lorigo et al., 2008; Pan et al., 2007) have determined which parts of the results pages are perceived by the users. Typical viewing patterns (such as the "golden triangle") have been determined and can be used to improve results presentation. Hotchkiss (2007) discussed the Google triangle, which clearly indicates users' focus on the upper left corner. However, this only holds true for a list-based results presentation, in which all results are presented equally. For unequal results presentations, quite different viewing patterns are found (see Figure 3).

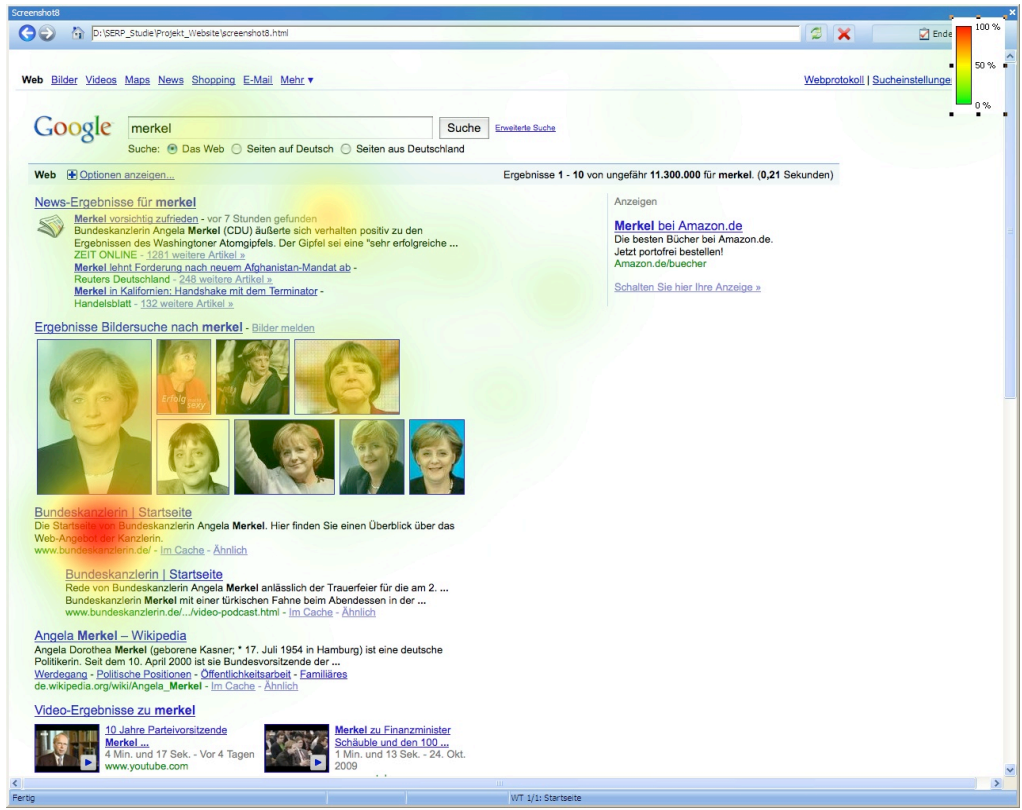

*Figure 3. Viewing pattern on a results page, revealing an unequal results presentation (heatmap from an eye-tracking study of 50 testing persons, in which the stimulus was displayed for 5 seconds).*

## A FRAMEWORK FOR EVALUATING THE RETRIEVAL EFFECTIVENESS OF SEARCH ENGINES

In the following sections, a framework for evaluating the retrieval effectiveness of search engines is presented. The framework consists of five parts, namely query selection, results collection, results weighting, results judgment, and data analysis. While different choices regarding the individual stages of the test design are made for different tests, guidance for designing such tests is given.

In the section pertaining to query selection, the kinds of queries that should be utilized when evaluating search engines are discussed. In the section on results collection, the information collected in addition to the URLS of the results are detailed. The results weighting section deals with the positions and the higher visibility of certain results, due to an emphasized presentation. The section on results judgment addresses who should make relevance judgments, what scales should be used, and how click-through data can support retrieval effectiveness tests. In the last portion of the framework (data analysis), appropriate measures that go beyond the traditional metrics of *recall* and *precision* are discussed. The framework, including all of the elements described below, is depicted in Figure 4.

# Framework for Evaluating the Retrieval Effectiveness of Search Engines

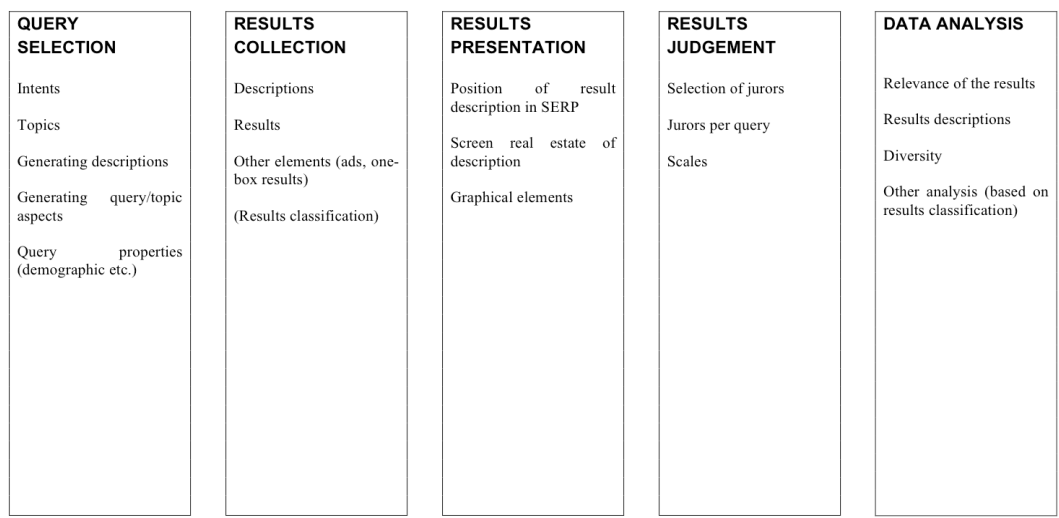

| QUERY SELECTION | RESULTS COLLECTION | RESULTS PRESENTATION | RESULTS JUDGEMENT | DATA ANALYSIS |
|---|---|---|---|---|
| Intents | Descriptions | Position of result description in SERP | Selection of jurors | Relevance of the results |
| Topics | Results | | Jurors per query | Results descriptions |
| Generating descriptions | Other elements (ads, one-box results) | Screen real estate of description | Scales | Diversity |
| Generating query/topic aspects | (Results classification) | Graphical elements | | Other analysis (based on results classification) |
| Query properties (demographic etc.) | | | | |

*Figure 4. Framework for the evaluation of the effectiveness of search engines.*

## Query selection

This section of the framework deals with selecting appropriate queries to evaluate. This selection is of great importance, as, due to limited resources, such tests must rely on a relatively small selection of queries. The use of queries with certain intent, query topics, how query descriptions (that is, descriptions of the underlying information needs) can be generated, how different aspects of a query can be taken into account, and how query properties can be used in evaluations are discussed below.

### Query intent

As already mentioned in the literature review section, queries used in testing should be separated according to type. In information science, a differentiation is made between a Concrete Information Need (CIN) and a Problem-Oriented Information Need (POIN) (Frants, Shapiro, & Voiskunskii, 1997). A CIN asks for factual information, and is satisfied with only one factum. In the case of document retrieval, this may mean satisfaction with only one document containing the required factum. In contrast, a POIN requires a smaller or larger number of documents for satisfaction.

Broder (2002) differentiates between three different types of intentions in terms of querying Web search engines: informational, navigational, and transactional. Navigational queries aim at a Web page that is already known to the user or to one which the user assumes exists (for example, the homepages of companies such as Ebay or people such as John von Neumann). Such queries normally terminate in one correct result. The information need is satisfied when the requested page is found.

In contrast, informational queries require more than one document (POIN). The user wishes to become informed about a topic and therefore intends to read several documents. Informational queries aim at static documents to acquire the desired information, which makes further interaction with the Web page unnecessary.

Transactional queries, however, aim at Web pages offering the possibility of a subsequent transaction such as the purchase of a product, the downloading of data, or the searching of a database.

Broders's taxonomy was used as a starting point by researchers who refined and amended it. (Kang & Kim, 2003) use classes very similar to Broder's, but use different notations ("topic relevance task" refers to an informational query, "homepage finding task" to a navigational query, and "service finding task" to a transactional query).

In terms of the evaluation of search engines, the different types must be considered separately (Lewandowski, 2008d, 2011a). Current studies mainly use informational queries or mix different query types. In tests using "real-life queries," the intent could (at least to some degree) be determined by using click-through data (see Joachims, 2002). The evaluation of search systems should always be oriented towards the queries that are put to this special type of search system.

It should be kept in mind that when a search engine's performance for two or more *different* query types is to be evaluated, the researcher must build distinct collections of queries and apply different metrics in the data analysis (see below).

## Query topics

Topics should reflect users' needs in the search engine under investigation. To achieve a statistically valid sample of queries according to their topics, one can classify a sample from the logfiles of the search engine under investigation (cf. Lewandowski, 2006).

It is important to use a variety of topics. However, when one wants to know how the search engine performs with different topics, sample sizes must be big enough to allow for such analysis.

## Generating descriptions

For an ideal retrieval effectiveness test, the juror should be the same person who initially posed the query to the search engine, as this person is the only one who can judge whether the results are appropriate for the fulfillment of his information need (Lewandowski, 2008d). However, in testing search engines, it is only seldom feasible to have users as jurors who use their own queries from real-life situations. As the queries themselves are, for the most part, not meaningful to others, it is indispensable to generate descriptions of the underlying information needs for each query. Ideally, a person describing her own query should do this.

However, when using queries from a logfile (which has the advantage that one can differentiate between popular and rare queries or even weight the queries by frequency), other methods must be used. One such method is to have different people describe what they think the underlying information need for a query is. Then, the different information needs are compared and similar information needs are synthesized. This leads to a basic set of information needs that can be used for the test. This method was used in (Huffman & Hochster, 2007), and produced good results.

## Generating query aspects

While the generation of query descriptions may reveal a variety of completely different information needs underlying the same query, documents answering the same query may cover more than one aspect of a query. For example, the query "James Bond" may involve aspects such as films, books, character biographies, actors who played the character, and so forth. While certain documents may contain general information about the topic and therefore cover many or even all aspects in brief, some documents may cover one or more aspects in depth. It is important in the evaluation of search engines to differentiate between topics and to honor a search engine that produces a result set that covers at least the major aspects on the first list of results presented to a user.

Query aspects can be generated in the same manner as the descriptions of the underlying information need.

## Query properties

There are far more query properties than query intents and query aspects. Probably the most comprehensive classification of query properties can be found in (Calderon-Benavides, Gonzalez-Caro, & Baeza-Yates, 2010). The authors describe nine facets that can be used to characterize a query. They consider genre, topic, task (which refers to Broder's classification and considers "informational," "not informational," and "both"), objective (where "action" is an indication of a commercial intent, and "resource" refers to a non-commercial intent), specificity, scope, authority sensitivity, spatial sensitivity, and time sensitivity.

Even demographic data can be obtained for the queries. Demographics Prediction, a free tool from Microsoft[ii], allows one to see whether a query is generally male- or female-oriented and what age groups pose this query at what frequency. While the data is for Web search, it can also be used for queries from other search engines. However, one should keep in mind that the demographics might be quite different between individual search tools.

Query descriptions, query aspects, and other query properties can be generated in one combined step. One can use jurors for generating these query properties or use a combined approach with jurors and the specific person who actually posed the query to the search engine (if this person is available).

## Results collection

The second pillar of the proposed framework deals with the collection of the results and additional information pertaining to the results from the search engines under investigation. Such additional information may include the results description, information on the type of result (direct answer vs. organic result description leading to a document), the position on the results screen, or the space a certain result obtains on the results screen.

### Results descriptions

The results descriptions are of great importance because they help a user to decide whether or not to select a particular result (Lewandowski, 2008d). However, the descriptions are not considered in the vast majority of evaluation studies. This distorts the results of retrieval effectiveness studies, in that results that would not have been clicked are considered. Research has shown that descriptions can be misleading. The titles and the descriptions of recommended Web pages are very important for users to evaluate the results lists (Machill, Neuberger, Schweiger, & Wirth, 2003; Machill et al., 2004). Descriptions should be collected in the same manner as the results themselves and should be considered in the results judgment, as well as in the data analysis phase.

### Capturing organic results

How results from a search engine should be captured depends on the nature of the engine(s) under investigation. If one is dealing with a system that searches a static database (for example, a test collection built for evaluating the search engine), there is no need to save the results, as they will not vary within the testing period. However, when using search engines not under the control of the investigator (whether because the search engine's database is not static or because changes may occur at any time during the testing period), all results must be saved in their entirety. While in the first case it may be enough to capture the results' URLs, in the second case the complete results must be saved.

The position of each individual result must also be captured. When comparing different search engines, the name of the search engine that produced an individual result must be captured.

### Capturing other elements of the results pages

When results are not presented equally, the type of result should be captured, as different result types may be perceived differently by the user. For example, the results page shown in Fig. 1 depicts results of organic, advertisement, shortcut, and extended snippet types (i.e., for an organic result, additional links are shown within the description). Fig. 2 shows a results description with an added picture.

### Results classification

Not only queries can be further classified, but also the results presented by a search engine. In Web searches, results may come from blogs, news sites, government agencies, and other types of pages. If the database searched by a search engine is not quality-controlled in a strict sense (that is, generally all types of documents from (a part of) the free Web are accepted for inclusion in the index), it may be useful to provide a good mix of results of different types.

In Web search, it may also be useful to classify results according to their commercial intent, as top results may be heavily influenced by search engine optimization (SEO) techniques (Lewandowski 2011b).

Results classification may reveal biases in results sets that were not intended in the design of a search engine. Additionally, different search engines may follow different approaches in mixing results of different types, which may affect the relevance of the *results set*, while pure relevance judgments may not provide an indication as to why a certain results set is considered of low relevance in its entirety.

## Results weighting

In this section, we consider results weighting, which involves weighting the relevance of results according to position and design. As we know from eye-tracking research and click-through analysis, results are not perceived equally. However, this is not considered in traditional retrieval effectiveness studies.

Results weighting can provide a weighting score for each results description, either according to its position on the search engine results page (SERP), the space it uses on the SERP, or the graphical accentuation it receives. For example, when considering the results list shown in figure 2, the result with the picture could receive a higher weighting score, as a user is more likely to perceive (and therefore, click) this result than the others that are simply presented on a text-basis.

### Position of the result descriptions within the SERP

While traditional retrieval effectiveness tests consider the results position (see also the section on data analysis below), this may not be enough when the results presentation is either (1) not list-based or (2) a list with more than one column. Consider the results presentation depicted in figure 1: the right-hand column presents context-based advertisements, which should also be considered as results, although they are surely of a different type. However, as can be seen from eye-tracking research and click-through analysis, these results are not perceived as strongly as the organic results in the "main column" of the results page. When results are collected for a retrieval effectiveness test, the position of the individual result within a result *list* should be captured, as well as its position within the results *page*.

### Screen real estate

"Screen real estate" is a term introduced by Nielsen & Tahir (2002) to measure the amount of space taken up by a Web page that is filled with content (vs. the part that is left blank). Nicholson et al. (2006) use screen real estate to calculate the space that search engines grant organic results vs. paid results on the first results screen shown to the user. While this is a specific application in the context given, screen real estate can be calculated for every result presented by a search engine. The use of screen real estate may seem obvious when results are presented in different formats; however, it can also be used for list-based results presentation, when not all results are equally presented. Figure 1 demonstrates that the first organic result is given far more real estate than the other results. Consequently, the first result should be given more weight when measuring retrieval effectiveness.

### Graphic elements

Search engine results pages do not necessarily consist of text-based results descriptions alone. Some results may be presented with additional pictures (as shown in Figure 3) or may be emphasized by the use of color. For example, the context-based ads depicted in Figure 1 (above the organic results list) are highlighted in yellow. Users thus perceive these results more, giving them a higher probability of being clicked.[iii] While such results could be weighted utilizing their screen real estate, it seems reasonable to ascribe even more weight to them due to their high influence on user behavior.

## Results judgment

This section discusses how jurors should be selected, how many queries a juror should be given to evaluate, and what scales should be used for relevance judgments. As these elements can influence the validity of retrieval effectiveness tests, choices should be made wisely.

### Selection of jurors

In most retrieval effectiveness studies, students are selected as jurors (see the literature review above), as they are easy to recruit and are usually willing to judge large numbers of documents, when relevance judgments are included as part of their coursework. However, while students may be good judges for general-purpose queries (as those used in Web search studies), they may not be good judges in other contexts, in which more background knowledge is needed. Consider, for example, a search engine for pharmaceutical information from the Web. Students (with the exception of students majoring in pharmacy) simply lack the background knowledge to evaluate the results appropriately. Studies using the "wrong" jurors are flawed; clearly, the effort of conducting a study should be avoided, when it is obvious that the results will be of limited use. Jurors should always be chosen according to the topic and target group of the search engine.

Another approach entails using a convenience sample of users to judge the results. This approach seems reasonable, as long as the target group of the search engine is homogeneous. When different user groups are targeted, more judgments per query should be selected (see below).

Jurors should clearly be selected with great care, as only a relatively small number of jurors can be recruited, due to the cost and effort retrieval effectiveness studies require.

### Jurors per query

Assuming that each juror reviews all of the results produced for a given query, we refer to "jurors per query" instead of "jurors per document." Although the latter terminology may seem more precise, it has the potential of leading to the false assumption that documents resulting from a single query are judged by different persons, that is to say, that a juror may judge only a part of the documents returned for an individual query.

Ideally, a large number of jurors per query should be used in a test so as to ensure that even when people judge results differently, these differences would be represented in the test. However, due to limited resources, most retrieval effectiveness studies utilize only one juror per query. When experimenting with a crowdsourcing-approach that allowed results to be judged by multiple jurors, it was determined that differences in results judgments are low and can therefore be neglected in a context that is familiar to the vast majority of users (Web search) and for more general queries,. In other contexts, on the other hand, more jurors per query should be used. Unfortunately, only a limited amount of research has been done on inter-rater agreement in relevance judgments (see Schaer, Mayr, & Mutschke, 2010). Therefore, a clear recommendation regarding the ideal number of jurors cannot yet be made. However, when a search engine targets different user groups, jurors from each group should judge the same documents, where applicable.

### Scales

When asking jurors to give their relevance judgments, it must be decided whether these judgments should be binary (relevant or not relevant) or collected using scales. While the first approach allows for ease in collecting data and computing precision scores (see "data analysis" below), the results may be flawed, as it is often not too difficult for a search engine to produce results that are *somehow* relevant, but can prove considerably more challenging to produce results that are *highly* relevant. Consequently, a more differentiated collection of relevance judgments seems preferential, although simple (e.g., 5-point) scales are to be preferred over, for example, more complex percent scales.

## Data analysis

In this section, the ways in which the data gathered in retrieval effectiveness tests can be analyzed are discussed, alongside the major measures used in such tests. Recommendations on what measures to use are also given.

### Relevance of the results

The foremost goal of retrieval tests is to measure the relevance of the results provided by the search engines under investigation. However, as relevance is a concept not clearly defined (Borlund, 2003; Hjorland, 2010; Mizzaro, 1997; Saracevic, 2007), an overwhelming number of relevance measures exist (for an overview, see Demartini & Mizzaro, 2006). In the remainder of this section, the "classic" relevance measures will be presented, as well as certain newer measures that seem promising for search engine evaluation purposes.

Retrieval performance of the IR system is usually measured by the "two classics," precision and recall. *Precision* measures the ability of an IR system to produce only relevant results. Precision is the ratio between the number of relevant documents retrieved by the system and the total number of documents retrieved. An ideal system would produce a precision score of 1; that is to say, every document retrieved by the system would be deemed relevant.

The other classic measure, *recall*, measures the ability of an IR system to find the complete set of relevant results from within a collection of documents. Recall refers to the ratio of the number of relevant documents retrieved by the system to the total number of relevant documents for the given query. As the total number of relevant documents must be known for each query within the database, recall is quite difficult to measure. Apart from retrieval effectiveness studies using test collections, recall cannot be calculated. A proposed solution to this problem is the method of pooling results from different engines and then measuring the relative recall of each engine.

Precision and recall are not mathematically dependent upon one another, but as a rule of thumb, the higher the precision of a results set, the lower the recall, and vice versa. For example, a system only retrieving one relevant result receives a precision score of 1, but usually scores a low recall. Another system that returns the complete database as a result (perhaps thousands or even millions of documents) will earn the highest recall but a very low precision.

Other "classic" retrieval measures include fallout and generality (for a good overview of retrieval measures see Korfhage, 1997). Newer approaches to measuring the quality of search results are as follows:

- *Median Measure* (Greisdorf & Spink, 2001), which takes into account the total number of results retrieved. Median measure measures not only how positive the given results are, but also how positive they are in relation to all negative results.
- *Importance of completeness of search results* and *Importance of precision of the search to the user* (Su, 1998). These two measures attempt to integrate typical user needs into the evaluation process. Whether the user requires a few precise results or a complete result set (while accepting a lower precision rate) is taken into account. These two measures seem highly promising for search engine evaluation purposes that focus on the user.
- *Value of Search Results as a Whole* (Su, 1998) is a measure that seems to correlate well with other retrieval measures that are regarded as important. Therefore, it can be used to shorten the evaluation process, making it less time-consuming and less costly.
- *Salience* refers to the sum of ratings for all hits for each service out of the sum of ratings for all services investigated (Ding & Marchionini, 1996). This measure takes into account how well all search engines studied perform on a certain query.
- *Relevance concentration* measures the number of items with ratings of 4 or 5 [on a five-point relevance scale] in the first 10 or 20 hits (Ding & Marchionini, 1996).
- *CBC ratio* (MacCall & Cleveland, 1999) measures the number of content-bearing clicks (CBC) in relation to the number of other clicks in the search process. A CBC can be defined as "any hypertext click that is used to retrieve possibly relevant information, as opposed to a hypertext click that is used for other reasons, such as the 'search' click that begins a database search or a 'navigation' click that is used to traverse a WWW-based information resource" (p. 764).
- *Quality of result ranking* takes into account the correlation between search engine ranking and human ranking (Vaughan, 2004, p. 681).
- *Ability to retrieve top ranked pages* combines the results retrieved by all search engines considered and lets them be ranked by humans. The "ability to retrieve top ranked pages" then measures the ratio of the top 75 percent of documents in the results list of a certain search engine (Vaughan, 2004).

As can be seen, retrieval measures vary considerably. Since the measure(s) chosen can strongly influence the results, the measures that are to be applied should be carefully selected, according to the goal of the search engine(s) under investigation. Results from analysis using different measures should be weighted against each other. As most of the retrieval measures proposed could be calculated from data collected by means of traditional retrieval measures, it is unproblematic to choose additional measures in the analysis of the data.

### Results descriptions

Results descriptions play an important role, as the user makes a judgment about whether or not to select a result based on the description. Since descriptions can sometimes be deceptive, they should be taken into account in all retrieval effectiveness tests. Even the best results are of no use to the user if he ignores them because of bad descriptions.

To clarify the underlying problem, an overview of the possible description/result pairs follows:

- *Relevant description* → *relevant result*: This is the ideal solution and should be provided for every result.

- *Relevant description → irrelevant result*: In this case, the user assumes from the description that the result is relevant and clicks on the results link, which leads the user to an irrelevant result. In most cases, the user returns to the search engine results page, but is frustrated to a certain degree.
- *Irrelevant description → irrelevant result*: The user generally does not click on the results link because of the irrelevant description and, therefore, does not examine the irrelevant result. One could say that such results descriptions are useful in that they at least warn the user not to click on the irrelevant results. However, these results descriptions really should not be presented on the results pages in the first place.
- *Irrelevant description → relevant result*: In this case, the user is unlikely to consider the result due to its misleading description. The user thus misses a relevant result on account of the poor description.

To measure the ability of search engines to produce consistent description/result pairs, we introduced specific measures (Lewandowski, 2008d):

- *Description-result precision* measures the ratio of results in which the description as well as the result itself were judged relevant. This measure can be regarded as measuring a kind of "super-precision;" every search engine should aim to provide only such results.
- *Description-result conformance* takes into account all pairs for which the description and result are judged the same (either both are deemed relevant or both are deemed irrelevant).
- *Description fallout* measures the ratio of results missed by a searcher due to a description that appears to point to an irrelevant result (that is, the description is judged irrelevant, while the actual result was judged relevant).
- *Description deception* measures in how many cases a user is led to an irrelevant result due to a description that appears to lead to a relevant result. Description deception can result in frustrated users and should be avoided by search engines.

### Diversity

Depending on the context, a diversity of sources is essential to obtaining a good result set. However, diversity within the result set may not be enough, as diversity should really be presented on the first result screen in order to give users an overview of different sources that have information on a particular topic. Consider, for example, a user seeking out information regarding minimum wage discussions. The first result screen should already present information from different sources. The user may be interested in the position of political parties, as well as the positions of the employers' federation and trade unions. Diversity of sources is relatively simple to measure, as the higher the number of sources presented on the first result screen, the better.

However, diversity of sources does not measure whether the query results cover all aspects of the underlying information need. When aspects are collected in the results collection phase (see above), these aspects can be extracted from the documents found and compared to the former. Thereafter, the number of results a user must view in a certain search engine until all aspects of the topic are displayed can be measured. Even if this appears to be too strict of an approach and it can be argued that users rarely require access to information on *all* different aspects of a topic, measuring aspects can be a good way of identifying search engines that produce relevant results, but where these results are too similar to each other.

### Other analysis (based on results classification)

If results classification was applied in the results collection step, further analysis can reveal whether certain types of results are more useful to the user than others. Consider, for example, results sets that feature results from weblogs. While in certain contexts (such as Web searches), these results may prove useful, in other contexts (for example, a search engine for trustworthy news), jurors may judge them as less useful. Separately calculating precision scores (or other retrieval measures) for the different kinds of results can reveal certain of these types to be more useful than others.

## CONCLUSIONS AND FUTURE RESEARCH DIRECTIONS

In this chapter, a new framework was presented for measuring the retrieval effectiveness of search engines. It is hoped that the framework proves useful in designing retrieval effectiveness tests.

While the work presented in this chapter is theoretical in nature, we plan to conduct extensive empirical studies using the framework. In the past, it has been difficult to utilize all of the elements of the framework for retrieval effectiveness tests, due to the great number of elements considered and the considerable effort that would result from employing them all. However, we have since developed a software tool meant to assist such tests, which makes it possible to conduct tests using the complete framework without inappropriate effort.

A major shortcoming of the proposed framework is that it fails to take user interaction into account. Information searching is more complex than the "query-response paradigm" (used in the framework) suggests (Marchionini, 2006). Search sessions are often quite a bit longer and include query modifications. Therefore, retrieval effectiveness studies would do well to consider search sessions as well, rather than merely focusing on isolated queries (Singer, Norbisrath, Vainikko, Kikkas, & Lewandowski, 2011). While considering search sessions could over-complicate such studies, the relationship of queries and sessions should at least be modeled (see Huffman & Hochster, 2007).

**ADDITIONAL READING**

## KEY TERMS & DEFINITIONS

### Query description

As queries themselves are often ambiguous and/or can mislead jurors, they must be further explained if the person judging the documents in a retrieval effectiveness test is not the same person who initially posed the query. Therefore, short descriptions of the queries used in the study are generated. The aim of the query description is to describe the underlying information need.

### Query type

Queries posed to search engines can be divided into certain query types. The most popular means of distinction is Andrei Broder's classification into informational, navigational, and transactional queries, where informational queries target a results set, navigational queries target a certain (known) item, and transactional queries target an interaction on the result.

### Results screen

A results screen refers to the portion of a search engine results page (SERP) that is visible to the user without scrolling. The size of the results screen can differ from one user to the next due to monitor and/or window size.

### Retrieval measure

A retrieval measure is used to quantify a certain aspect of the performance of a search engine. Most measures used in retrieval effectiveness studies deal with the perceived relevance of the documents presented by a certain search engine. However, some retrieval measures properties of the results presented (such as the ratio of a certain document type) into account.

### Search engine results page (SERP)

A search engine results page is a complete presentation of search engine results; that is, it presents a certain number of results (determined by the search engine). To obtain more results, a user must select the "further results" button, which leads to another SERP.

### Screen real estate

The screen real estate refers to the area on the results screen that is given a certain element (e.g., a result description). When results are presented unequally, screen real estate can be used to measure the relative importance of a certain result within the search engine results page or the results screen.

---

[i] "TREC-style" refers to the test design used in the TREC ad-hoc tracks.

[ii] http://adlab.msn.com/Demographics-Prediction/

[iii] This does not consider that a certain number of Web search engine users developed a pattern of ignoring the ads because they know that these results are advertisements. The example can be regarded as an illustration of how user perception can be directed by emphasizing certain results.